\documentclass[a4paper]{JHEP3}
\usepackage{amsmath,amsfonts,a4,epsfig,graphicx}
\usepackage{multicol}
\usepackage{float}
\numberwithin{equation}{section}
\def\nn{\nonumber}

\def\beq{\begin{equation}}
\def\eeq{\end{equation}}
\def\bea{\begin{eqnarray}}
\def\eea{\end{eqnarray}}
\def\lsim{\mathrel{\raisebox{-.6ex}{$\stackrel{\textstyle<}{\sim}$}}}
\def\gsim{\mathrel{\raisebox{-.6ex}{$\stackrel{\textstyle>}{\sim}$}}}

\catcode`\@=11

\def\@citex[#1]#2{\if@filesw\immediate\write\@auxout{\string\citation{#2}}\fi
  \def\@citea{}\@cite{\@for\@citeb:=#2\do
    {\@citea\def\@citea{,\penalty\@m}\@ifundefined
       {b@\@citeb}{{\bf ?}\@warning
       {Citation `\@citeb' on page \thepage \space undefined}}%
\hbox{\csname b@\@citeb\endcsname}}}{#1}}

\def\citer{\@ifnextchar
[{\@tempswatrue\@citexr}{\@tempswafalse\@citexr[]}}
\def\@citexr[#1]#2{\if@filesw\immediate\write\@auxout{\string\citation{#2}}\fi
  \def\@citea{}\@cite{\@for\@citeb:=#2\do
    {\@citea\def\@citea{--\penalty\@m}\@ifundefined
       {b@\@citeb}{{\bf ?}\@warning
       {Citation `\@citeb' on page \thepage \space undefined}}%
\hbox{\csname b@\@citeb\endcsname}}}{#1}}
\catcode`\@=12

\preprint{PSI--PR--10--11}

\title{Higgs Boson Production via Gluon Fusion in the Standard Model
with four Generations}

\author{Qiang Li$^{\, (1)}$, Michael Spira$^{\, (1)}$ \\
$^{(1)}$ Paul Scherrer Institut, CH--5232 Villigen PSI, Switzerland  
}

\author{Jun Gao$^{\, (2)}$, Chong Sheng Li$^{\, (2),(3)}$ \\
$^{(2)}$ Department of Physics and State Key
Laboratory of Nuclear Physics and Technology, Peking
University, Beijing, 100871, China \\
$^{(3)}$ Center for High Energy Physics, Peking
University, Beijing, 100871, China \\

}

\abstract{Higgs bosons can be produced copiously at the LHC via gluon
fusion induced by top and bottom quark loops, and can be enhanced
strongly if extra heavy quarks exist. We present results for Higgs +
zero-, one- and two-jet production at the LHC operating
at 7 and 14 TeV collision energy, in both the Standard
Model and the 4th generation model, by evaluating the corresponding
heavy quark triangle, box and pentagon Feynman diagrams. We compare the
results by using the effective Higgs-gluon interactions in the limit of
heavy quarks with the cross sections including the full mass
dependences. NLO effects on Higgs + zero-jet production rate with full mass dependence
are presented for the first time consistently in the 4th generation model.
Our results improve the theoretical basis for fourth generation effects
on the Higgs boson search at the LHC.}

\date{\Date}


\begin{document}


\section{Introduction}
\label{intr}

The Large Hadron Collider (LHC) is presently running with a c.m.~energy
of 7 TeV. One of its main goals is to explore the details of electroweak
symmetry breaking, in particular the Higgs sector of the Standard Model
(SM).  At the LHC Higgs boson production via gluon fusion $gg\to H$
\citer{GF,GFHQexamples} is the dominant production mode. More exclusive
channels such as Higgs associated production with one or two hard jets
via gluon fusion or weak boson fusion
\citer{deFlorian:1999zd,Campanario:2010mi} have also been studied. By
using the accompanying jet information, one can refine the experimental
cuts to increase the signal to background ratio and the potential of
extracting Higgs parameters.

The measurement of Higgs boson production at the LHC can also give hints
or constraints on those new physics beyond the SM which couple to the
Higgs sector, such as e.g.~the fourth generation extension of the SM
(SM4) (see for example
\citer{Frampton:1999xi,Erler:2010sk}), which assumes additional heavy
quarks ($b^\prime$, $t^\prime$) with Higgs Yukawa interactions growing
with their masses as for the SM quarks. The SM4 can address some of the
current open questions, e.g.~it provides new sources for CP violation
\cite{Hou:2008xd} and baryogenesis \cite{Carena:2004ha,Kikukawa:2009mu}.
The SM4 is also consistent with the precision electroweak data. In
particular the SM4 fit can lead to a much higher upper limit on the
Higgs mass of $\sim 750$\,GeV at 95\% C.L.~\cite{He:2001tp,Kribs:2007nz} than the
SM one, and thus reduces the tension with the 114.4 GeV lower limit from
LEP2 \cite{LEPII,PDG}. Recently the Tevatron constrained the
4th-generation $b'$ mass to
$m_{b^\prime}>338$\,GeV~\cite{Aaltonen:2009nr}, while unitarity
requirements bound the 4th generation $t'$ mass as
$m_{t^\prime}<504$\,GeV~\cite{Erler:2010sk,unitarityc}. Notice also that
a SM4 Higgs with mass between 131 GeV and 204 GeV has been excluded at 95\% C.L.
by the Tevatron~\cite{Aaltonen:2010sv}.

In the SM4, the appearance of heavy flavours enlarges the Higgs coupling
to gluons significantly and thus the Higgs boson production rate at
hadron colliders. Refs.~\cite{Anastasiou:2009kn,Anastasiou:2010bt}
present the NNLO QCD results for Higgs inclusive production with
effective Higgs-gluon interactions, and further incorporates the
contributions of the top and extra heavy quarks consistently following
the method of Ref.~\cite{Anastasiou:2008tj}. In
Ref.~\cite{Kribs:2007nz}, Higgs+2-jet production via gluon fusion has
been studied as a background to the weak boson fusion process by using
the effective Higgs coupling to gluons. In this paper, we
investigate Higgs+jet(s) associated production, and check in particular
the validity of the approach of using the effective Higgs-gluon
interactions in our case. This paper is organized as follows. In
Section~\ref{calculations} we describe the calculation. In
Section~\ref{results} we present numerical results and their
discussion. Finally we conclude in Section~\ref{sec:end}.

\section{Description of the Calculation}
\label{calculations}

At the LHC, Higgs+1-jet and 2-jet production via gluon fusion receives
contributions from the partonic processes
\begin{eqnarray}\label{Dia1}
gg\rightarrow gh\, ,\quad gq\rightarrow qh\, ,\quad q\bar{q}\rightarrow
gh ,
\end{eqnarray}
and
\begin{eqnarray}\label{Dia2}
  & &gg\rightarrow ggh, \, q\bar{q}h , \nn\\
  & &gq\rightarrow gqh, \nn \\
  & &qq\rightarrow qqh,\quad qq'\rightarrow qq'h,\nn\\
  & &q\bar{q}\rightarrow ggh,\,  q\bar{q}h, \,  q'\bar{q}'h,
\end{eqnarray}
respectively, where $q'$ denotes quarks with different flavour than $q$.

The relevant one-loop Feynman diagrams and amplitudes for all the
subprocesses mentioned above have been generated with FeynArts 3.5
\cite{FeynArts}, and manipulated with FormCalc 5.3 \cite{FormCalc}.  The
Fortran libraries generated with FormCalc are then linked with our Monte
Carlo integration code for final use, in which we have modified the
codes generated with FormCalc by adding the extra heavy flavour
contributions of the SM4. The tensor integrals are evaluated with the
help of the LoopTools-2.5 package \cite{FormCalc}, which employs the
reduction method introduced in Ref.~\cite{PentagonA} for pentagon
tensor, and Passarino-Veltman reduction for the lower point ones. The
resulting regular scalar integrals are evaluated with the FF
package~\cite{FF}. Note that the UV and IR divergent scalar integrals
have already been encoded into this newest version of LoopTools within
dimensional regularization, which we have checked with QCDloop
\cite{QCDloop}. Moreover, we have modified LoopTools-2.5 to implement
the reduction method for pentagon tensor integrals as proposed in
Ref.~\cite{PentagonB}, and found it leads to much better numerical
stability in our case.

We have performed a second calculation based on the heavy-top
effective Lagrangian
\cite{GFNLOexamples,Shifman:1979eb},
which is a good approximation for not too heavy Higgs bosons ($M_h\lsim
m_t$) and in appropriate kinematic regions \cite{DelDuca:2001eu}:
\begin{equation}
  \label{higgs1}
  \mathcal{L}_{eff}=n_h\frac{\alpha_s}{12\pi v}H
  G_{\mu\nu}^aG^{a\mu\nu},
\end{equation}
where $G^a_{\mu\nu}$ denotes the gluon field strength tensor, and $n_h$
represents the number of heavy quarks, i.e. $n_h=1 (3)$ for the SM
(SM4). This effective model has already been implemented in MadGraph4
\cite{Alwall:2007st}, with which we generated all contributing tree
level Feynman diagrams and helicity amplitudes for the processes listed
in Eqs.~(\ref{Dia1},\ref{Dia2}).  The numerical evaluation is then
performed by using the HELAS library \cite{helas}.

We have checked our calculations in several ways. First, we compared the
results of our two calculations and got good agreement between them for
smaller Higgs masses as expected. Second, we compared the SM result with
Ref.~\cite{DelDuca:2001eu} and could reproduce their results with the
same settings and parameter choices.

\section{Numerical Results}
\label{results}

In this section we present the total cross sections and differential
distributions for Higgs+1-jet and 2-jet production
at the LHC. For completeness, we also show the relevant Higgs+0-jet
results at LO and NLO.

We impose the minimal set of cuts
\begin{eqnarray}\label{cut1}
&& |\eta_{j}|< 4.5 \,,\,\qquad P_T^{j}>P_T^{\, {\rm cut}}\,,\, \qquad
\Delta R_{jj}= \sqrt{\Delta \eta^2+\Delta \phi^2}> 0.6\;
\end{eqnarray}
to identify massless partons with jets. Here $\eta$ is the
pseudorapidity of the jets and $\phi$ is the azimuthal angle around the
beam direction. $P_T^{\, {\rm cut}}$ is the jet transverse momentum cut,
chosen as a function of the Higgs mass $M_h$:
\begin{eqnarray}\label{cut1b}
P_T^{\, {\rm cut}}=0.04\, M_h+14\,{\rm GeV},
\end{eqnarray}
thus for example $P_T^{\, {\rm cut}}=30$\,GeV for
$M_h=400$\,GeV, which ensures the perturbatively
reliability of our results over a wide range of Higgs mass, i.e.
$\sigma_{2j}<\sigma_{1j}<\sigma_{0j}$.

We also show the results for Higgs+2-jet production with the following
weak-boson-fusion cuts in addition,
\begin{eqnarray}\label{cut2}
|\eta_{j1}-\eta_{j2}|>4, \qquad \eta_{j1}\cdot\eta_{j2}<0, \qquad
m_{jj}>600\;{\rm GeV}.
\end{eqnarray}

Throughout our calculation, we set the top quark mass to
$m_t=173.0$\,GeV and take the five flavour scheme to treat external
bottom quarks as massless particles while keeping the bottom quark mass
as $m_b=4.6$\,GeV in the fermion loops. The Fermi constant has been
chosen as $G_F=1.16637\times 10^{-5}$ GeV$^{-2}$. For the LO results we
adopt the CTEQ6L1 parton distribution functions (PDFs)
\cite{Pumplin:2002vw} with the corresponding value for the LO strong
coupling $\alpha_s(M_Z)=0.130$, while at NLO we used the CTEQ6.6M PDFs
with the NLO strong coupling normalized to $\alpha_s(M_Z)=0.118$.  Our
default choice for the renormalization and factorization scales is
$M_h$.

For the SM4 parameters, we first focus on the scenario as
chosen in Ref.~\cite{Anastasiou:2010bt}:
\begin{eqnarray}\label{SM4parameter}
m_{b^\prime}=400\,{\rm GeV},
\end{eqnarray}
and
\begin{eqnarray}\label{mtprime}
m_{t^\prime}=m_{b^\prime}+50\,{\rm
GeV}+10\log\bigg(\frac{M_h}{115{\rm GeV}}\bigg){\rm GeV},
\end{eqnarray}
which is consistent with electroweak precision tests
\cite{Kribs:2007nz}. We will also discuss the dependence on
$m_{b^\prime}$.

\FIGURE{\label{pphnj}
\includegraphics[width=12.cm]{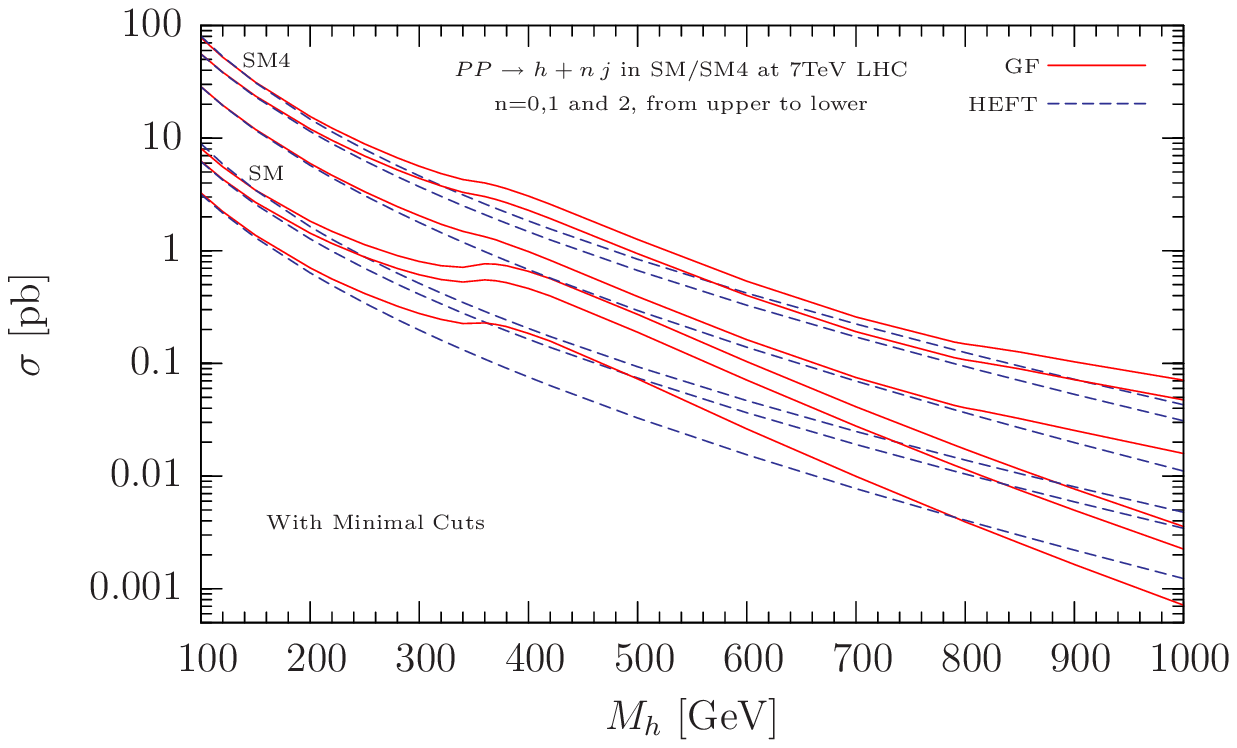}
\includegraphics[width=12.cm]{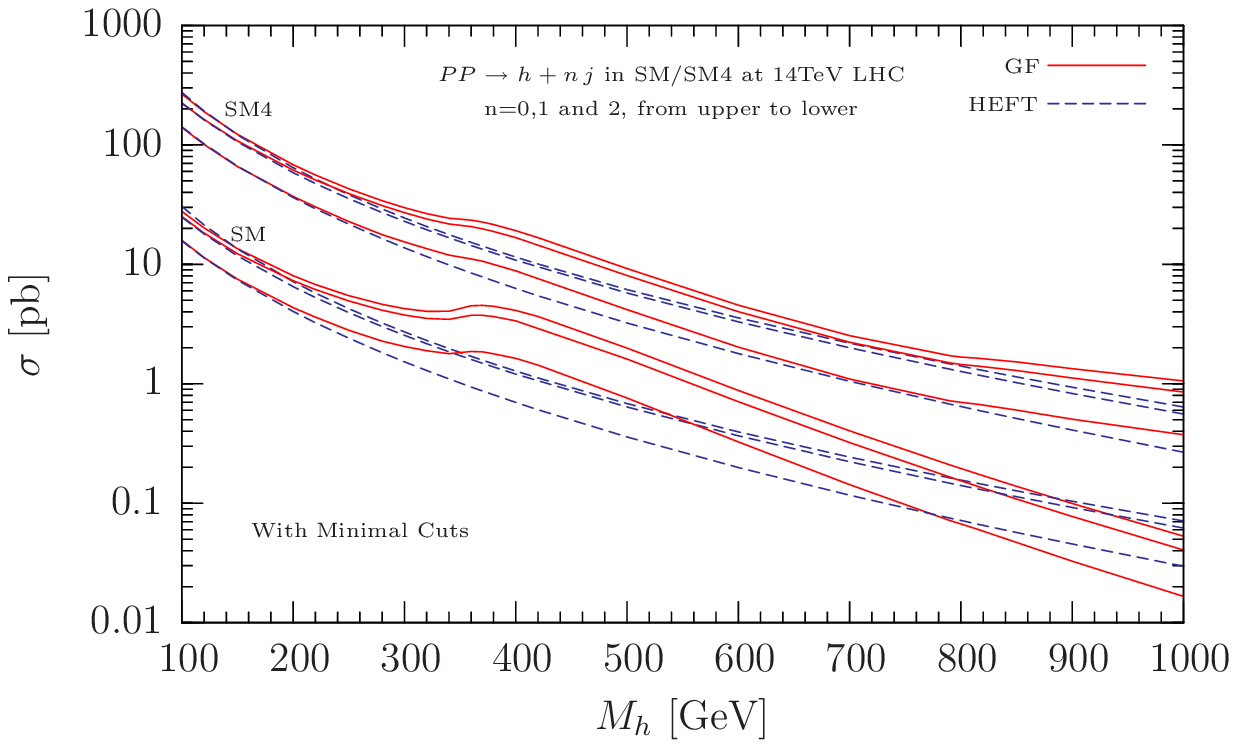}
\caption{Cross sections of Higgs+$n$-jet ($n=0,1,2$) production via
gluon fusion (GF) in the SM and SM4 ($m_{b^\prime}=400$\,GeV) at the LHC with
$\sqrt{s}=7$ and $14$ TeV as functions of the Higgs boson mass with the minimal
cuts of Eq.~(\ref{cut1}). The corresponding effective Higgs couplings
results (HEFT) are also shown.}}

Fig.~\ref{pphnj} shows the Higgs+$n$-jet ($n=0,1,2$) production cross
sections as functions of the Higgs boson mass at the LHC with
$\sqrt{s}=7$ and $14$ TeV including the minimal cuts of Eq.~(\ref{cut1}). As
expected, in the light Higgs boson region ($M_h \lsim 200$\,GeV) the
gluon-fusion results agree well with the ones obtained by using the
effective Higgs coupling to gluons for both the SM and SM4. The visible
discrepancy for the $n=0$ case in the light Higgs boson region is due to
the contribution of the bottom quark loops at the 10\%-level, which is
much smaller for Higgs+1-jet and 2-jet production. For light Higgs
bosons one can approximate the enhancement ratio of the SM4 rate over
the SM one as $n_h^2\approx 9$. In the larger Higgs mass region, the
effective Higgs coupling approximation breaks down, and the gluon-fusion
results differ significantly from the effective Higgs coupling
approximation, especially when $M_h$ is near the thresholds at $M_h\sim
2m_{t,b^\prime,t^\prime}$, where threshold effects play a role.  For
heavy Higgs bosons with $M_h\gsim 800$\,GeV the Higgs+1-jet and 2-jet
production cross sections at the LHC with
$\sqrt{s}=14$ ($7$) TeV in the SM4 can exceed the SM one by more than
a factor of 10 and amount to ${\cal O}(100)~fb$ (${\cal O}(10)~fb$), which is promising for
the related Higgs boson searches at the LHC.

\FIGURE{\label{pphjjwbf}
\includegraphics[width=12.cm]{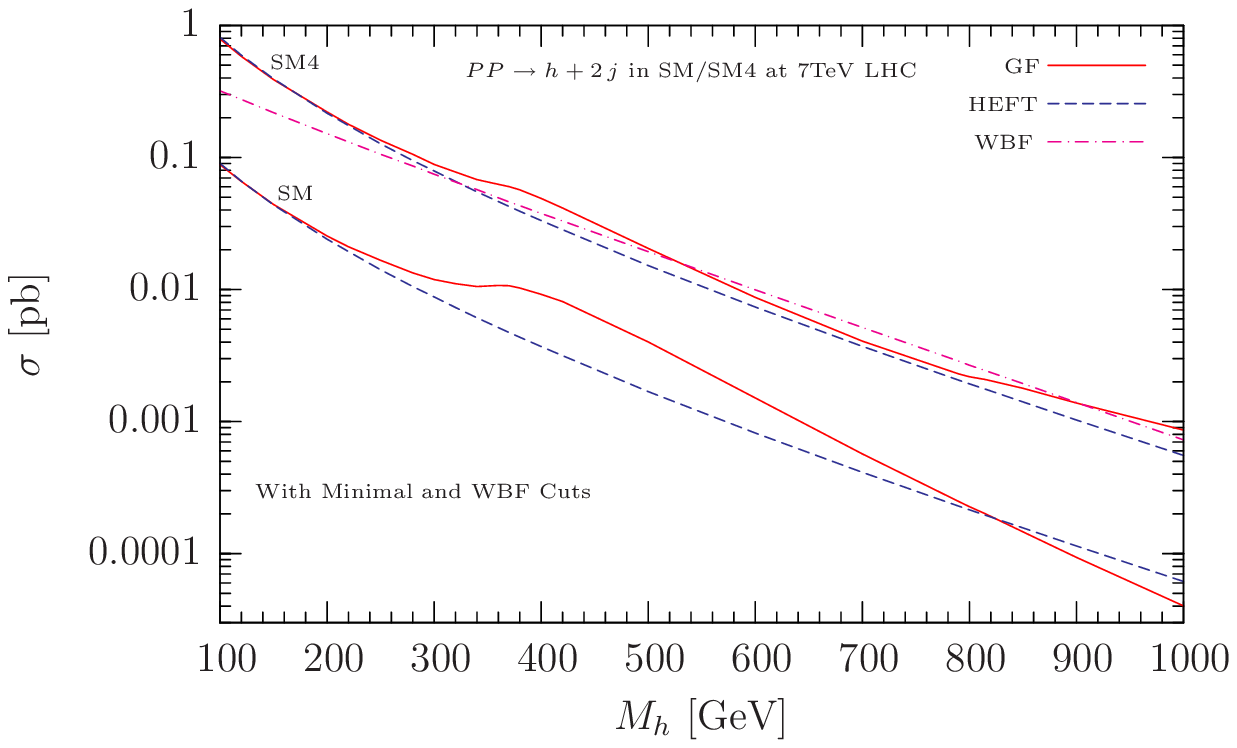}
\includegraphics[width=12.cm]{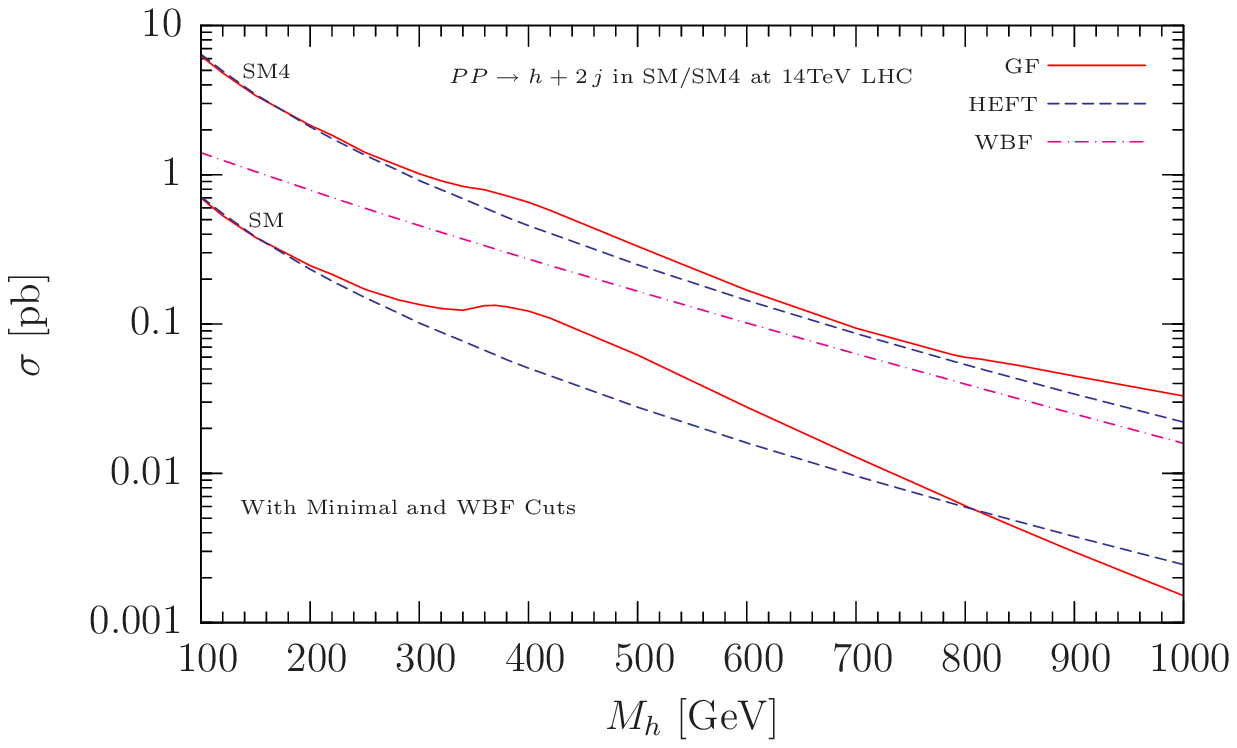}
\caption{Cross sections of Higgs+2-jet production via gluon fusion (GF)
in the SM and SM4 ($m_{b^\prime}=400$\,GeV) and via weak boson fusion
(WBF) in the SM at the LHC with $\sqrt{s}=7$ and $14$ TeV as functions of the
Higgs boson mass with both the minimal and weak-boson-fusion cuts of
Eqs.~(\ref{cut1},\ref{cut2}). The corresponding effective Higgs coupling
results (HEFT) for the gluon fusion processes are also shown.}}

In Fig.~\ref{pphjjwbf}, we show the cross sections as functions of $M_h$
for Higgs+2-jet production via gluon fusion in both the SM and SM4, and
also display the SM results via the weak-boson-fusion process, imposing
both the minimal and weak-boson-fusion cuts of
Eqs.~(\ref{cut1},\, \ref{cut2}). Again, the effective Higgs coupling
approximation works well for Higgs masses below about 200\,GeV, for both
the SM and SM4, even after imposing the large jet invariant mass cut in
Eqs.~(\ref{cut2}), in agreement with Ref.~\cite{DelDuca:2001fn}. One
observes that the SM4 results are much larger than the SM ones, and
can exceed the weak-boson-fusion results to a large extent, especially at the 14TeV LHC. Thus the
importance of the weak-boson-fusion channels will be diminished and the
Higgs search strategies will be affected within the SM4 context. 

Fig.~\ref{ptmax} displays the transverse momentum distribution of the
harder jet for Higgs+2-jet production via gluon fusion in the SM and SM4
at the 14TeV LHC including the minimal cuts of Eq.~(\ref{cut1}) for
$M_h=125$\,GeV and $400$\,GeV, respectively. For both the SM and SM4,
the effective Higgs coupling approximation works well if the Higgs mass
and the $P_T$ of the harder jet is small ($P_T\lsim m_t$), as found in
Ref.~\cite{DelDuca:2001fn}. The agreement is better in the SM4 because
the contributions of the heavier extra quarks dominate and thus postpone
the breakdown of the approximation to larger Higgs boson mass values.
However, the transverse momentum distribution of the harder jet is
softer than in the effective Higgs coupling approximation in general.

\FIGURE{\label{ptmax}
\includegraphics[width=11.cm]{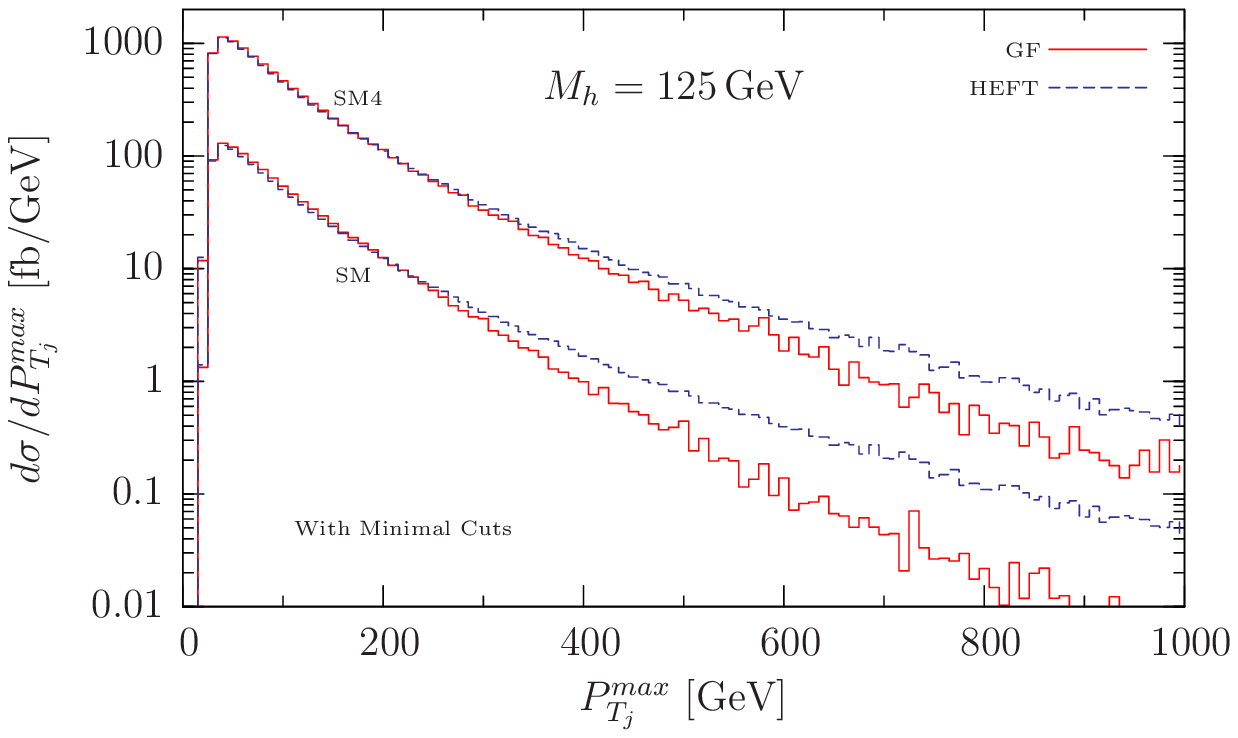} \\ \hspace*{-1.2cm}
\includegraphics[width=11.cm]{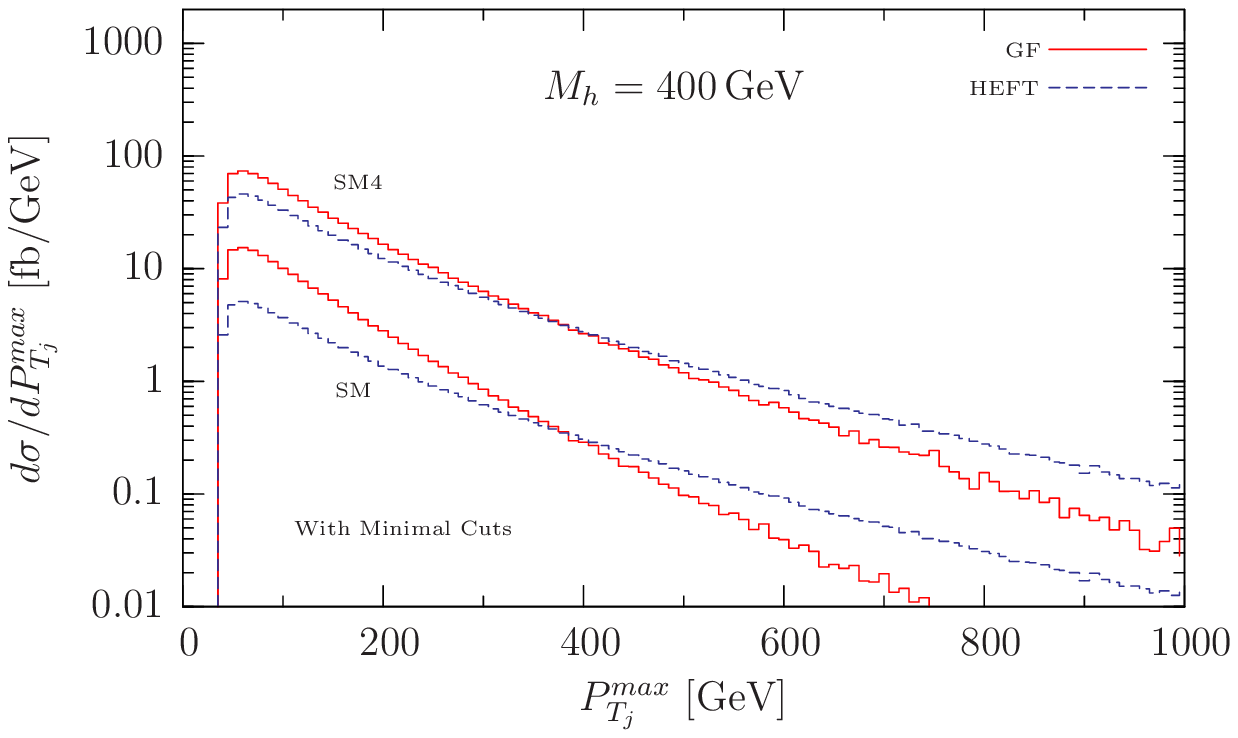}
\caption{Transverse momentum distribution of the harder jet for
Higgs+2-jet production via gluon fusion (GF) in the SM and the SM4
($m_{b^\prime}=400$\,GeV) at the LHC with $\sqrt{s}=14$ TeV for
$M_h=125$\,GeV (left) and $400$\,GeV (right). The minimal cuts of
Eq.~(\ref{cut1}) are imposed. The corresponding effective Higgs
coupling results (HEFT) are displayed as the blue dashed histograms.}}

\FIGURE{\label{phi}
\includegraphics[width=11.cm]{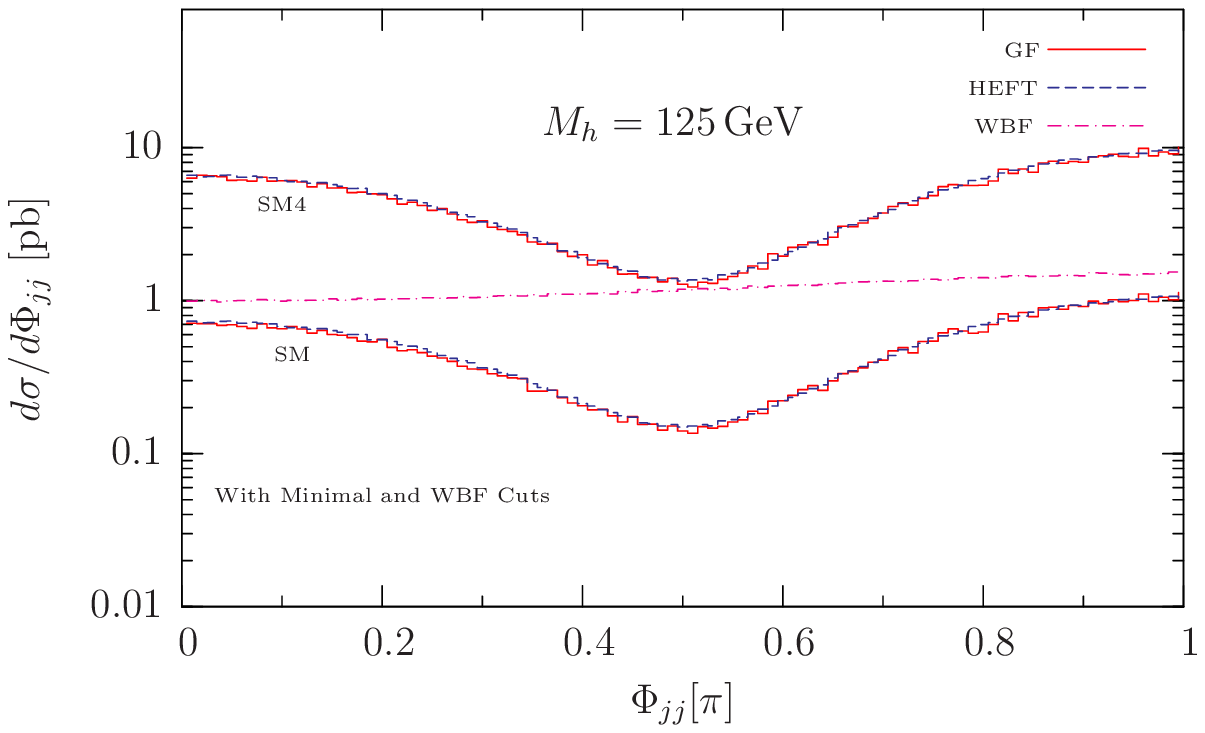} \\ \hspace*{-1.2cm}
\includegraphics[width=11.cm]{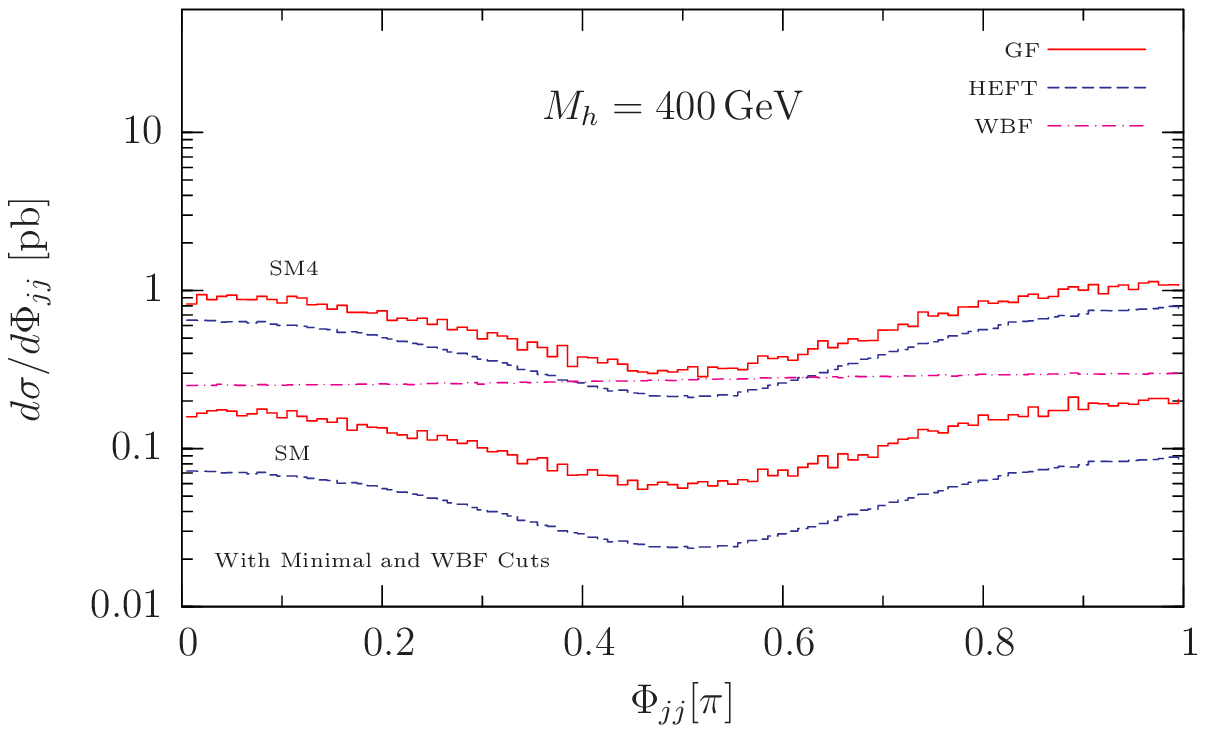}
\caption{Azimuthal angle distribution between the two jets for
Higgs+2-jet production via gluon fusion (GF) in both the SM and SM4
($m_{b^\prime}=400$\,GeV), and via weak boson fusion (WBF) in the SM at
the LHC with $\sqrt{s}=14$ TeV for $M_h=125$\,GeV (left) and $400$\,GeV
(right). Both the minimal and weak-boson-fusion cuts of
Eqs.~(\ref{cut1},\ref{cut2}) have been imposed. The corresponding
effective Higgs coupling results (HEFT) are also shown.}}

In Fig.~\ref{phi}, we show the azimuthal angle distribution between the
two jets for Higgs+2-jet production via gluon fusion in both the SM and
SM4 ($m_{b^\prime}=400$\,GeV) for $M_h=125$\,GeV and $400$\,GeV,
respectively. We also display the corresponding gluon-fusion results
with the effective Higgs coupling approximation, and the
weak-boson-fusion ones within the SM. Again the gluon fusion and
effective Higgs coupling results agree well for the whole $\Phi_{jj}$
region for the light Higgs boson case ($M_h=125$\,GeV) but not for the
heavy one ($M_h=400$\,GeV).  One can observe the characteristic flat
distribution of the weak-boson-fusion and the dip structures of the
gluon fusion results, respectively
\citer{Plehn:2001nj,Hagiwara:2009wt} for both the SM and SM4, which
originate from the properties of the CP-even Higgs boson couplings to
weak bosons and to gluons within the SM. The SM4 results are larger than
the SM ones, and are close to the weak-boson-fusion results even in the
central region $\Phi_{jj}\sim 0$, which will affect the Higgs boson
search via the weak-boson-fusion channel significantly within the SM4.

In Tables~\ref{sublist}-\ref{sublist2} we present the cross sections for Higgs+$n$-jet
($n=0,1,2$) production via gluon fusion at the 7TeV and 14TeV LHC in the SM and SM4
with the minimal cuts of Eq.~(\ref{cut1}) and $m_{b^\prime}=300, 400$
and $500$ GeV, respectively.  The NLO numbers for Higgs+0-jet production
have been obtained from HIGLU~\cite{Spira:1995mt} by vetoing jets within the cuts of
Eq.~(\ref{cut1})\footnote{It should be noted that the analysis of Ref.~\cite{Arik:2005ed}
used several approximations
so that the full mass effects have not been taken into account consistently at NLO.}.
We also exhibit the scale variation uncertainties by varying the
renormalization and factorization scales $\mu_R=\mu_F$ from $M_h/2$ to
$2M_h$\footnote{Note that the scale dependences at LO are expected to
underestimate the real theoretical uncertainties but serve as a rough
estimate. Due to the large QCD corrections to Higgs+1-jet and 2 jet
production known in the SM in the heavy top limit
\cite{deFlorian:1999zd,Campbell:2006xx} one expects that the full QCD
corrections within the SM4 will be sizeable and positive in general.}.
With increasing $m_{b^\prime}$ the cross sections in the SM4 become
smaller by $\lsim 50\%$ for $m_{b^\prime}$ varied from $300\,{\rm GeV}$
to $500\,{\rm GeV}$. The scale dependence of Higgs+2-jet
(1-jet) production is much larger than for Higgs+1-jet (0-jet)
production at LO due to the additional jet emission. The NLO numbers for
Higgs+0-jet production indicate moderate QCD corrections and a strong
reduction of the residual scale dependence. We expect a similar strong
reduction of the scale dependence for Higgs+1-jet and 2-jet production
at NLO after vetoing additional jets within the minimal cuts of
Eq.~(\ref{cut1}). The strong uncertainty for Higgs+jet production at LO
may hide the effects of extra heavy quarks partly in future LHC
searches. However, due to the large enhancement of the SM4 compared to
the SM production rates and the accompanying jet information, the
Higgs+jet production channels can still be quite useful for early
discovery of a 4th generation within the SM context. Moreover, 
motivated by Refs.~\cite{Catani:2001ve,Krauss:2002ly,Hoeche:qf,Alwall:2007bh},
we are presently performing the merging of the above calculated LO matrix elements
of various jet multiplicities with parton showers in a consistent way, to reach
fully exclusive description of events
for Higgs production at the LHC in both the SM and SM4~\cite{alm}.

\TABLE{\label{sublist}\centering{
\begin{tabular}{|c|c|c|c|c|} \hline
7TeV LHC    & SM & SM4-300 & SM4-400 & SM4-500 \\ \hline
$0j$ LO & $17.4^{+6.5}_{-4.4}$ & $248^{+92}_{-63}$ & $150^{+56}_{-38}$
& $117^{+44}_{-30}$ \\ \hline
$0j$ NLO  & $27.7^{+1.6}_{-2.8}$  &  $423^{+36}_{-49}$ & $254^{+21}_{-29}$
& $194^{+14}_{-21}$ \\ \hline
$1j$ & $11.5^{+5.8}_{-3.6}$ & $170^{+85}_{-53}$ & $108^{+54}_{-34}$
& $86.7^{+43.2}_{-26.9}$ \\  \hline
$2j$ & $3.97^{+2.5}_{-1.5}$  & $60.8^{+39}_{-22}$ &
$40.7^{+26}_{-15}$ & $33.1^{+21}_{-12}$ \\ \hline
\end{tabular}
\caption{Dependence of the Higgs+$n$-jet ($n=0,1,2$) production cross
sections (in fb) via gluon fusion on the heavy quark masses and their
scale uncertainty (by varying the factorization and renormalization
scales by a factor of 2 up- and downwards) at the 7TeV LHC with the minimal
cuts of Eq.~(\ref{cut1}) for $M_h=800$\,GeV.}}}

\TABLE{\label{sublist2}\centering{
\begin{tabular}{|c|c|c|c|c|} \hline
14TeV LHC & SM & SM4-300 & SM4-400 & SM4-500\\ \hline
$0j$ LO & $195.5^{+60.8}_{-43.5}$ & $2776^{+864}_{-618}$ & $1677^{+522}_{-373}$
& $1316^{+409}_{-293}$ \\ \hline
$0j$ NLO & $227.2^{+1.4}_{-11.6}$  & $3506^{+76}_{-239}$  & $2107^{+41}_{-142}$
& $1597^{+10}_{-96}$ \\ \hline
$1j$ & $154.2^{+68.1}_{-44.2}$ & $2273^{+1003}_{-651}$ & $1452^{+641}_{-416}$
& $1163^{+514}_{-333}$ \\  \hline
$2j$ & $68.0^{+39.8}_{-23.5}$  & $1043^{+611}_{-361}$ &
$705^{+413}_{-244}$ & $577^{+338}_{-200}$ \\ \hline
\end{tabular}
\caption{Same as Table~\ref{sublist}, but for 14TeV LHC.}}}

For the experimental Higgs search within the SM4 the large modifications
of the Higgs branching ratios may be highly relevant in addition. For light Higgs ($M_h<131$\,GeV),
in analogy to the gluon-fusion processes the decay width of the Higgs boson
into gluons will be strongly enhanced, thus reducing all other branching
ratios accordingly. Moreover, the additional heavy quark contributions
will diminish the $W$-loop contribution to $H\to\gamma\gamma$, so that
this rare decay mode will be suppressed even further. Due to these
changes of the Higgs boson profile within the SM4 context, the search
strategies have to be reinvestigated in detail. For heavier Higgs ($M_h>204$\,GeV), 
the decay channels $H\rightarrow ZZ, W^+W^-$ or $t\bar{t}$ remain the most 
dominate ones and the relevant branch ratios are nearly not changed, except near the threshold region 
(2$m_{b^\prime, t^\prime}$), where new decay channels of Higgs into 4th generation quarks are opened.
Higgs boson searches via the weak-boson-fusion processes will be affected in a similar way by the
modified Higgs branching ratios.

\section{Summary}
\label{sec:end}
We have presented the calculations of Higgs+1-jet and 2-jet production
processes induced by gluon fusion at the LHC, in both the SM and its
fourth generation extention.  We have compared the full results to the
corresponding ones by using the effective Higgs couplings to gluons in
the heavy quark approximation. As in
Refs.~\cite{DelDuca:2001eu,DelDuca:2001fn}, we have found that the
approximation works well in both the SM and SM4, for light Higgs bosons
and moderate transverse momenta of the jets, $P_T\lsim m_t$, while
otherwise the differences are large in general. Light Higgs boson
production within the SM4 with jet(s) via gluon fusion is approximately
enhanced by a factor 9 compared to the SM cross sections. For heavy
Higgs bosons the production rates amount to more than 10 times the SM
one and reach values of ${\cal O}(100~fb)$.  However, these results are
plagued by significant scale uncertainties, but they can still be useful
for early discovery of the Higgs boson with extra heavy quark
contributions. A fourth generation will affect the Higgs search strategy
via the weak boson fusion channel strongly.

\acknowledgments
This work is supported in part by the Concerted Research action
Supersymmetric Models and their Signatures at the Large Hadron Collider
of the Vrije Universiteit Brussels, by the IISN "MadGraph" convention
4.4511.10, by the National Natural Science Foundation of China, under
Grants No. 10721063, No. 10975004 and No. 10635030 and the European
Community's Marie-Curie Research Training Network HEPTOOLS under
contract MRTN-CT-2006-035505.


\appendix



\begin{thebibliography}{00}

\bibitem{GF}
H.~M.~Georgi, S.~L.~Glashow, M.~E.~Machacek and D.~V.~Nanopoulos,
Phys.\ Rev.\ Lett.\ {\bf 40}, 11 (1978)

\bibitem{GFNLOexamples}
S.~Dawson, Nucl.\ Phys.\ B {\bf 359}, 283 (1991);
A.~Djouadi, M.~Spira and P.~Zerwas, Phys.\ Lett.\ B {\bf 264}, 440 (1991);
D.~Graudenz, M.~Spira and P.~Zerwas, Phys.\ Rev.\ Lett. {\bf 70}, 1372 (1993);
M.~Spira, A.~Djouadi, D.~Graudenz and P.~Zerwas, Nucl.\ Phys.\ B {\bf 453}, 17 (1995).

\bibitem{GFHQexamples}
R.~V.~Harlander and W.~B.~Kilgore, Phys.\ Rev.\ Lett.\  {\bf 88} (2002) 201801;
C.~Anastasiou and K.~Melnikov, Nucl.\ Phys.\  B {\bf 646} (2002) 220;
V.~Ravindran, J.~Smith and W.~L.~van Neerven, Nucl.\ Phys.\  B {\bf 665} (2003) 325;
M.~Kr\"amer, E.~Laenen and M.~Spira, Nucl.\ Phys.\  B {\bf 511} (1998) 523;
S.~Catani, D.~de Florian, M.~Grazzini and P.~Nason, JHEP {\bf 0307} (2003) 028;
S.~Moch and A.~Vogt,
  Phys.\ Lett.\ B {\bf 631} (2005) 48
  [arXiv:hep-ph/0508265];
V.~Ravindran,
  Nucl.\ Phys.\ B {\bf 746} (2006) 58
  [arXiv:hep-ph/0512249];

\bibitem{deFlorian:1999zd}
  D.~de Florian, M.~Grazzini and Z.~Kunszt,
  Phys.\ Rev.\ Lett.\  {\bf 82}, 5209 (1999)
  [arXiv:hep-ph/9902483].

\bibitem{Mellado:2007fb}
  B.~Mellado, W.~Quayle and S.~L.~Wu,
  Phys.\ Rev.\  D {\bf 76}, 093007 (2007)
  [arXiv:0708.2507 [hep-ph]];
  U.~Langenegger, M.~Spira, A.~Starodumov {\it et al.},
  JHEP {\bf 0606 } (2006)  035.
  [hep-ph/0604156];
  O.~Brein and W.~Hollik,
  arXiv:0710.4781 [hep-ph].

\bibitem{Rainwater:1997dg}
  D.~L.~Rainwater and D.~Zeppenfeld,
  JHEP {\bf 9712}, 005 (1997)
  [arXiv:hep-ph/9712271];
  D.~L.~Rainwater, D.~Zeppenfeld and K.~Hagiwara,
  Phys.\ Rev.\  D {\bf 59}, 014037 (1999)
  [arXiv:hep-ph/9808468].

\bibitem{Plehn:2001nj}
  T.~Plehn, D.~L.~Rainwater and D.~Zeppenfeld,
  Phys.\ Rev.\ Lett.\  {\bf 88}, 051801 (2002)
  [arXiv:hep-ph/0105325].

\bibitem{DelDuca:2001eu}
  V.~Del Duca, W.~Kilgore, C.~Oleari, C.~Schmidt and D.~Zeppenfeld,
  Phys.\ Rev.\ Lett.\  {\bf 87}, 122001 (2001)
  [arXiv:hep-ph/0105129].

\bibitem{DelDuca:2001fn}
  V.~Del Duca, W.~Kilgore, C.~Oleari, C.~Schmidt and D.~Zeppenfeld,
  Nucl.\ Phys.\  B {\bf 616}, 367 (2001)
  [arXiv:hep-ph/0108030].

\bibitem{Campbell:2006xx}
John M.~Campbell, R.~Keith Ellis, Giulia Zanderighi,
JHEP {\bf 0610} (2006) 028.

\bibitem{Hankele:2006ja}
  V.~Hankele, G.~Klamke and D.~Zeppenfeld,
  arXiv:hep-ph/0605117;
  G.~Klamke and D.~Zeppenfeld,
  JHEP {\bf 0704}, 052 (2007)
  [arXiv:hep-ph/0703202].

\bibitem{Hagiwara:2009wt}
  K.~Hagiwara, Q.~Li and K.~Mawatari,
  JHEP {\bf 0907} (2009) 101
  [arXiv:0905.4314 [hep-ph]].

\bibitem{Campanario:2010mi}
  F.~Campanario, M.~Kubocz and D.~Zeppenfeld,
  arXiv:1011.3819 [hep-ph].

\bibitem{Frampton:1999xi}
  P.~H.~Frampton, P.~Q.~Hung and M.~Sher,
  Phys.\ Rept.\  {\bf 330}, 263 (2000)
  [arXiv:hep-ph/9903387].

\bibitem{He:2001tp}
  H.~J.~He, N.~Polonsky and S.~f.~Su,
  Phys.\ Rev.\  D {\bf 64}, 053004 (2001)
  [arXiv:hep-ph/0102144].

\bibitem{Kribs:2007nz}
  G.~D.~Kribs, T.~Plehn, M.~Spannowsky and T.~M.~P.~Tait,
  Phys.\ Rev.\  D {\bf 76}, 075016 (2007)
  [arXiv:0706.3718 [hep-ph]].

\bibitem{Holdom:2009rf}
  B.~Holdom, W.~S.~Hou, T.~Hurth, M.~L.~Mangano, S.~Sultansoy and G.~Unel,
  PMC Phys.\  A {\bf 3}, 4 (2009)
  [arXiv:0904.4698 [hep-ph]].

\bibitem{Erler:2010sk}
  J.~Erler and P.~Langacker,
  Phys.\ Rev.\ Lett.\  {\bf 105}, 031801 (2010)
  [arXiv:1003.3211 [hep-ph]].

\bibitem{Hou:2008xd}
  W.~S.~Hou,
  Chin.\ J.\ Phys.\  {\bf 47}, 134 (2009)
  [arXiv:0803.1234 [hep-ph]].

\bibitem{Carena:2004ha}
  M.~S.~Carena, A.~Megevand, M.~Quiros and C.~E.~M.~Wagner,
  Nucl.\ Phys.\  B {\bf 716}, 319 (2005)
  [arXiv:hep-ph/0410352].

\bibitem{Kikukawa:2009mu}
  Y.~Kikukawa, M.~Kohda and J.~Yasuda,
  Prog.\ Theor.\ Phys.\  {\bf 122}, 401 (2009)
  [arXiv:0901.1962 [hep-ph]].

\bibitem{LEPII}
LEP Electroweak Working Group, status of August 2009,
http://lepewwg.web.cern.ch/LEPEWWG/
\bibitem{PDG}
K.~Nakamura et al. (Particle Data Group), J.\ Phys.\ G {\bf 37},
075021 (2010).

\bibitem{Aaltonen:2009nr}
  T.~Aaltonen {\it et al.}  [CDF Collaboration],
  Phys.\ Rev.\ Lett.\  {\bf 104}, 091801 (2010)
  [arXiv:0912.1057 [hep-ex]].

\bibitem{unitarityc}
W.~J.~Marciano, G.~Valencia, and S.~Willenbrock, Phys.\ Rev.\ D {\bf
40}, 1725 (1989).

\bibitem{Aaltonen:2010sv}
  T.~Aaltonen {\it et al.}  [CDF and D0 Collaboration],
  arXiv:1005.3216 [hep-ex].

\bibitem{Anastasiou:2009kn}
  C.~Anastasiou, S.~Bucherer and Z.~Kunszt,
  JHEP {\bf 0910}, 068 (2009)
  [arXiv:0907.2362 [hep-ph]].

\bibitem{Anastasiou:2010bt}
  C.~Anastasiou, R.~Boughezal and E.~Furlan,
  JHEP {\bf 1006}, 101 (2010)
  [arXiv:1003.4677 [hep-ph]].

\bibitem{Anastasiou:2008tj}
  C.~Anastasiou, R.~Boughezal and F.~Petriello,
  JHEP {\bf 0904}, 003 (2009)
  [arXiv:0811.3458 [hep-ph]].

\bibitem{FeynArts}
J.~K\"ublbeck, M.~B\"ohm, and A.~Denner,
Comput. Phys.  Commun.  {\bf 60} (1990)  165--180;
  T.~Hahn,
  Comput.\ Phys.\ Commun.\  {\bf 140}, 418 (2001)
  [arXiv:hep-ph/0012260].
\bibitem{FormCalc}
  T.~Hahn and M.~Perez-Victoria,
  Comput.\ Phys.\ Commun.\  {\bf 118}, 153 (1999)
  [arXiv:hep-ph/9807565].
\bibitem{PentagonA}
  A.~Denner and S.~Dittmaier,
  Nucl.\ Phys.\  B {\bf 658}, 175 (2003)
  [arXiv:hep-ph/0212259].
\bibitem{FF}
G.~J.~van Oldenborgh and J.~A.~M.~Vermaseren, Z.\ Phys.\ C {\bf 46},
425 (1990), G.~J.~van Oldenborgh, Comput.\ Phys.\ Commun. {\bf 66}
(1991).
\bibitem{QCDloop}
  R.~K.~Ellis and G.~Zanderighi,
  JHEP {\bf 0802}, 002 (2008)
  [arXiv:0712.1851 [hep-ph]].
\bibitem{PentagonB}
  A.~Denner and S.~Dittmaier,
  Nucl.\ Phys.\  B {\bf 734}, 62 (2006)
  [arXiv:hep-ph/0509141].
\bibitem{Shifman:1979eb}
J.\ Ellis, M.K.\ Gaillard and D.V.\ Nanopoulos, Nucl.\ Phys.\
{\bf B106} (1976) 292;
M.~A. Shifman, A.~I. Vainshtein, M.~B. Voloshin, and V.~I. Zakharov,
Sov.\ J.\ Nucl.\ Phys. {\bf 30}, 711 (1979);
B.~A. Kniehl and M.~Spira, Z.\ Phys.\ C {\bf 69}, 77 (1995).
\bibitem{Alwall:2007st}
  J.~Alwall {\it et al.},
  JHEP {\bf 0709}, 028 (2007)
  [arXiv:0706.2334 [hep-ph]].
\bibitem{helas}
  H.~Murayama, I.~Watanabe and K.~Hagiwara,
  KEK-Report 91-11, 1992.

\bibitem{Pumplin:2002vw}
  J.~Pumplin, D.~R.~Stump, J.~Huston, H.~L.~Lai, P.~Nadolsky and W.~K.~Tung,
  ``New generation of parton distributions with uncertainties from global  QCD
  analysis,''
  JHEP {\bf 0207}, 012 (2002)
  [arXiv:hep-ph/0201195].

\bibitem{Spira:1995mt}
  M.~Spira,
  arXiv:hep-ph/9510347.


\bibitem{Arik:2005ed}
  E.~Arik, O.~Cakir, S.~A.~Cetin and S.~Sultansoy,
  Acta Phys.\ Polon.\  B {\bf 37}, 2839 (2006)
  [arXiv:hep-ph/0502050].

\bibitem{Catani:2001ve}
S.~Catani, F.~Krauss, R.~Kuhn, and B.~Webber, {\it {QCD Matrix Elements +
  Parton Showers}},  {\em JHEP} {\bf 0111} (2001) 063,
  [\href{http://xxx.lanl.gov/abs/hep-ph/0109231}{{\tt hep-ph/0109231}}].

\bibitem{Krauss:2002ly}
F.~Krauss, {\it Matrix elements and parton showers in hadronic interactions},
  {\em JHEP} {\bf 0208} (2002) 015,
  [\href{http://xxx.lanl.gov/abs/hep-ph/0205283}{{\tt hep-ph/0205283}}].

\bibitem{Hoeche:qf}
S.~Hoeche {\em et.~al.}, {\it {Matching parton showers and matrix elements}},
  \href{http://xxx.lanl.gov/abs/hep-ph/0602031}{{\tt hep-ph/0602031}}.

\bibitem{Alwall:2007bh}
J.~Alwall {\em et.~al.}, {\it {Comparative study of various algorithms for the
  merging of parton showers and matrix elements in hadronic collisions}},  {\em
  Eur. Phys. J.} {\bf C53} (2008) 473--500,
  [\href{http://xxx.lanl.gov/abs/0706.2569}{{\tt arXiv:0706.2569}}].

\bibitem{alm}
J.~Alwall, Q.~Li and F.~Maltoni, {\it {Matched predictions for Higgs production via heavy-quark loops in the SM and beyond }}, arXiv:1110.1728.

\end{thebibliography}
\end{document}